
\magnification=1200
\baselineskip=20pt
\hsize=15.0true cm
\hoffset=1.3true cm
\centerline{{\bf Did the Pseudo-Sphere Universe have a Beginning?}}
\vskip 0.3in
\centerline{G.Oliveira-Neto \footnote\dag{Email
address: G.OLIVEIRA-NETO@NEWCASTLE.AC.UK}}
\vskip 0.2in
\centerline{Department of Physics}
\centerline{University of Newcast1e-Upon-Tyne}
\centerline{Newcastle-Upon-Tyne,\quad NE1 \quad 7RU}
\centerline{U.K.}
\vskip 0.5in
\centerline{{\bf Pseudo-Spherical Universes}}
\vskip 0.5in
\centerline{ Classification Number - 460}
\vskip 0.3in
\centerline{ABSTRACT}
\vskip 0.2in
\par\noindent A calculation of the no-boundary wave-function of the universe
is put forward for a spacetime with negative curvature.  A semi-classical
Robertson-Walker approximation is attempted and two solutions to the
field equations, one Lorentzian and the other a tunneling one are found.
The regularity of those solutions are analysed explicitly, both in $2+1$
and $3+1$ dimensions and a conical singularity is found at the origin of
the time axis, contradicting the no-boundary assumption.
\vskip 0.3in
{\bf $1$\quad-\quad Motivation\quad and\quad Introduction}

          During the last few years, some effort has been spent on
understanding
models of the universe with negative spacetime curvature[1]. One can think of
at
least two main reasons, if the pleasure derived from the geometrical insights
furnished by those models is not enough: The low energy limit  of Supergravity
with  its  negative  energy vacuum state [2] and Wheeler's idea of a foam like
structure for the universe  at the Planck scale[3], with all possible
topologies
contributing  to spacetime.

         Our  motivation has to do with the second reason. We are interested in
building  up  wave-functions  for  universes  with  negative  curvature  in  a
consistent way, such that we could in the future  use them to  describe  a
more   complete picture, which would take in account  contributions  coming
from general topologies. Many examples  of  consistent  wave-functions for
space-times  with  positive  curvature have  already  been examined  [4].
Consistency  of the wave-function involves the definition of sensible initial
conditions, and  so  far the most appealing proposal is the one devised  by
Hartle \& Hawking [5], the so called no-boundary proposal.

         What we ultimately shall be doing here is trying to derive the
no-boundary wave-function for a space-time with negative curvature, where we
foliate the  spacetime  in  spatial sections which evolve in time  [6].   We
shall not try to derive general properties of these wave-functions, rather we
shall choose a simple example and  work  out  the  particular  wave-function
arising from this restricted choice.  For  mathematical  simplicity, and  also
to account for the present observational data we shall  choose our  model  to
be homogeneous and isotropic, which leads us into a minisuperspace
approximation. So, our metric ansatz will be Robertson-Walker, such that  the
spatial sections will have negative curvature, and our unique source of mass-
energy will be a cosmological constant, negative in this case.  Besides  that,
we shall be  concerned  with  the  semi-classical  approximation  of  the
no-boundary wave-function. In this task we shall have to  overcome  two  basic
problems:\hfil\break
(1)It  is  well  know  that  the Hartle \& Hawking proposal  demands  that  all
spacetimes contributing to the wave-function have compact spatial sections. But
one's intuition does not deal easily with the idea  of  a  compact  surface
with negative curvature [6].\hfil\break
(2) For spacetimes with negative curvature, the signature of solutions is
likely
to be Lorentzian, whereas the no-boundary proposal prefers Riemannian
signature.

Before we proceed any further we should ask what is known already. A couple of
years ago, Gibbons \& Hartle [7]  devised some general statements about the
predictions of the no-boundary wave-function for the present large-scale
topology and geometry of the universe. There, they argued that among the
complex
solutions of the Euclidean  Einstein's equations, important in the
semi-classical  limit,  the purely  real tunneling  ones play a special role.
The simplest example of those  solutions occurs in a model like ours, differing
by the opposite  sign  of the cosmological constant. In this model, the real
tunneling solution  (which corresponds to discontinuos metrics) is given by
joining of Riemannian metric with a Lorentzian one along a surface
characterized by the  vanishing of its second fundamental form.  One important
point is that  the  Riemannian region is related to the early universe, in
the  sense  that  the  scale  factor assumes in this sector its smallest
values, and grows up until one well defined value, which is dependent on the
unique physical parameter in this problem, the cosmological constant.  On the
other hand, the Lorentzian  region, has the property of starting off with the
final scale factor from the foregoing region and increases up to the value
assigned by the final boundary condition. The schematic picture below (borrowed
from them), gives a better idea of what was said above.
\vskip 0.4in
\topinsert
\vskip 2.0in
\centerline{1-{\sl Tunneling Solution}}
\endinsert
\vskip 0.3in
The question is: Can we take this result for granted in the case  of negatively
curved spacetimes? The  authors  mentioned  above [1], assumed the above result
also valid  for  spacetimes with negative curvature, but it will be argued
otherwise here.

In the next section, we shall derive the no-boundary solutions  for the simple
model. In section 3, with the help of few  mathematical  results, we shall be
able to see which are the compact  spatial  sections  allowed  in  this case.
In section 4,  we  shall  determine if these solutions are regular or not.
\vskip 0.3in
{\bf $2$\quad-\quad Euclidean\quad Negatively\quad Curved\quad
Robertson-Walker \hfill\break Spacetime}
\vskip 0.2in
         The general expression for the no-boundary wave-function for a space-
time [8]is simplified by the assumptions that it is homogeneous, isotropic and
its unique source of mass-energy is a negative cosmological constant. This
leads
to the Robertson-Walker metric ansatz, with the lapse function $N$ and the
scale
factor $a$  depending only on $t$ and the spatial sections beeing
three-dimensional surfaces with negative curvature ( we could have chosen to
foliate our four manifold in a different way, but one's intuition is led to
the choice below): \vskip 0.2in
$$ds^2=+N^2(t)dt^2+a^2(t)[d\chi^2+\sinh^2\chi(d\theta^2+\sin^2\theta
d\phi^2)]\eqno (1)$$
\vskip 0.2in
\par\noindent With this expression for the metric, the wave-function reduces in
complexity hugely to the more workable form, (in the gauge which $N$ is
constant):
\vskip 0.2in
$$\Psi_{nb}(\tilde a)= \int dN \int D[a(t)] e^{-I[a(t),N]}\eqno(2)$$
\vskip 0.2in
\par\noindent where $I[a(t),N]$ is the Euclidean action for the various
different geometries  contributing through the integrals of $N$ and $a(t)$ to
the wave-function. It is given by:
\vskip 0.2in
$$I[N(t),a(t)]={1\over 16\pi}\int (R-2\Lambda)\sqrt{\,+g}\,dx^4+{1\over
8\pi}\int K\sqrt{\,+h}\,dx^3\eqno (3)$$
\vskip 0.2in
\par\noindent in this expression we set the velocity of light and the
gravitational constant to unity, $R$ is the scalar curvature, $K$ is the the
trace of the extrinsic  curvature of the spacelike boundary surface,
$\Lambda$ is the cosmological  constant, $g$ is the trace of the metric and $h$
is the trace of the metric on the three-dimensional spatial sections. From the
metric ansatz (1), the action (3), can be straightforwardly computed,
\vskip 0.2in
$$I[N,a(t)]=Z\int[\dot a^2(t)-N^2-a^2(t)N^2{\Lambda\over 3}]a(t){dt \over N}
\eqno (4)$$
\vskip 0.2in
\par\noindent where $Z$ is a finite defined number, once one specifies the
compact spatial  slices which will foliate the whole manifold.

        All of the instructions needed in order to write down the
semi-classical
approximation of the no-boundary wave-function (2) for a given spacetime, of
which ours is a simple case, are given by Halliwell \& Louko [9]. The first
step we shall have to make to get the wave-function in the desired
approximation, which is given explicitly below, is to find the solutions to the
Euclidean-Einstein's equations. Then, \vskip 0.2in
$$\Psi_{nb}[a]=\sum_kP_k(a)e^{-I_k(a)}\eqno (5)$$
\vskip 0.2in
\par\noindent Where, $I_k$ are the euclidean actions of the solutions  of
the  Euclidean-Einstein's equations for the given metric ansatz. $P_k$  is
the  prefactor  and  k labels the various possible classical solutions over
which  we  must  sum  in order to obtain the correct wave-function and a is
the scale factor on the  final boundary surface.

          The pair of independent equations generated by the above metric
ansatz
(1), coming from the  Euclidean-Einstein  equations, with a negative
cosmological constant are:
\vskip 0.2in
 $$2\dot a a + \dot a^2-N^2(1-{a^2\over\vert\Lambda\vert})=0\eqno (6.1)$$

$$\dot a^2+N^2(1-\vert\Lambda\vert{a^2 \over 3})=0\eqno (6.2)$$
\vskip 0.2in
          Now, in possession of the general solutions, we must specify that
they
satisfy the 'no-boundary' boundary conditions. For our case [9]:\hfil\break
(1)The manifold must be regular at every point;\hfil\break
(2) There should be a point where the scale factor vanishes;\hfil\break
(3) We should supply the real valued scale factor of the final spatial
slice.\hfil\break The first condition we shall tackle after the other two are
implemented. The second one is a particularity of the minisuperspace treatment,
and we shall choose the point to be the zero value of the time scale. With this
choice we shall be able to picture our universe starting at t=0, from this
surface of zero volume and evolving until $t=t_1$, where we shall furnish the
other required value of the scale factor, say, $a(t_1)= a_1$.  Introducing
those
quantities in the above (6.1) and (6.2) equations, and rescaling the time scale
in order to write $t_1$ equal to $1$, we get the two solutions given below:
\vskip 0.2in
\par\noindent The Lorentzian Solution:

$$N=iN_I,\quad \hbox {where} \quad N_I= \alpha{{\pi\over 2}\pm
\arccos[{1\over \alpha}a_1]}\eqno (7.1)$$

$$a(t)=\pm\alpha\sin[N_I{t\over \alpha}]\qquad;\qquad
\alpha^2={3\over \vert\Lambda\vert}\eqno (7.2)$$
\vskip 0.2in
\par\noindent valid for $a_1 < \alpha$. It is important to notice that among
the
various solutions for $N$, labeled by an integer $m$, we have chosen as a
matter
of simplicity, the case $m=0$ (the same remark holds for the complex solution
given below).  The resulting metric is:
\vskip 0.2in
$$ds^2=-N_I^2dt^2+\alpha^2\sin[{1\over
\alpha}N_It]^2[d\chi^2+\sinh^2\chi(d\theta^2+\sin^2\theta d\phi^2)]\eqno
(7.3)$$
\vskip 0.2in
\par\noindent The  complex  solution:

$$N=\alpha(N_R + iN_I),\qquad \hbox {where} \qquad N_I={\pi \over 2}\qquad
\hbox {and}$$

$$N_R=\arcsin\beta \quad ;\quad \beta=\mp \sqrt[({a_1\over \alpha})^2-1]\eqno
(8.1)$$

$$a(t)=a_R(t)+ia_I(t)\quad ;\qquad  \hbox {where}$$

$$a_R(t)=\mp \alpha \sin(N_It)\cosh(N_Rt)\qquad \hbox {and}$$

$$a_I(t)=\pm \alpha \cos(N_It)\sinh(N_Rt)\eqno (8.2)$$
\vskip 0.2in
\par\noindent and for this case the metric is a complex one, as written below:
\vskip 0.2in
$$ds^2=\alpha^2(N_R^2-N_I^2)dt^2+(a_R^2(t)-a_I^2(t))\bar \Omega_3 +
2i\alpha^2(N_RN_Idt^2+a_R(t)a_I(t)\bar \Omega_3)\eqno (8.3)$$
\vskip 0.2in
\par\noindent where $\bar \Omega_3$ stands for the metric  of  the
three-dimensional  pseudo-sphere, and this solution holds for
$a_1 > \alpha$.

          The natural sequence of our  investigation,  would  be  to  address
the question of regularity, as demanded by the condition (1) above, to the two
solutions just found, in order for those  to  be  regarded  as  truly
no-boundary  solutions of our model. It is important to  notice,  that  these
solutions  are   well defined for all $t$, except $t=0$. There, the metrics
vanish identically, and our analysis of regularity must be focused on this
specific point. But  before  we begin we will try to write our complex solution
in the form of a real tunneling  one.

          As one can see, from the expression of the semi-classical no-boundary
wave-function, the quantities which dominate the wave-function are  the
classical actions, not the metrics themselves. It may  be  possible  to find
another metric for which, the new action is a composition of two  actions: the
first derived from a Lorentzian metric (which will give the purely imaginary
contribution to the action) and the second from a Riemannian  metric  (
responsible for the real part). Fortunately, for our simple model, it is easy
to  show  that such an interpretation of the complex solution is applicable.

          We should start by inspecting our action (4),  following
the  path of Halliwell \& Hartle [10]. It is easy to see that (4) can be
rewritten  as an integral over a complex variable, $T$, defined as the  product
$t\times N$. In  other  words,
\vskip 0.2in
$$I[a(T)]=Z\int[({da(T)\over dT})^2-1-a^2(T){\Lambda\over 3}]a(T)dT\eqno (9)$$
\vskip 0.2in
\par\noindent Now, we want to deform the contour of integration on  the
complex
$T$  plane, such that our contour breaks off in two parts (in this simple
case):
one  running along or parallel to the real axis (what would be interpreted as
the Riemannian sector), and another running along or parallel to the imaginary
$T$ axis (interpreted as the Lorentzian sector). One is left with two choices
of
contour, which are easily determined from the value of the coordinate $T$ at
the
final boundary surface. The value of the on this surface is:
\vskip 0.2in
$$\bar T_f =\bar T_{Rf} + i\bar T_{If} \qquad  \hbox {where},$$

$$\bar T_{Rf}=\alpha\arcsin\beta \quad ;\quad \bar T_{If}=\alpha{\pi \over
2}\eqno (10)$$
\vskip 0.2in
\par\noindent Then, our two possible contour are given in the diagram below:
\vskip 2.5in
\centerline {2-{\sl Deformations of contour on the complex $\bar T$ plane}}
\vskip 0.3in
      The circuit (a) above, was extensively studied  in  reference  [10]. The
condition one has to satisfy in order to have  a  purely  imaginary  action,
when our variable $T$ runs along the second part of this circuit (ii), is that
the first derivative of the scale factor at the point where this part of the
circuit starts, $P_a$ in the diagram above, must vanish (this is  the condition
that  the extrinsic curvature vanishes at this point, for our simple case).
{}From the explicit value of $a(T)$, in terms of this complex variable:
\vskip 0.2in
$$a(T) = -i\alpha\sinh({T\over \alpha})\;; \qquad
\hbox {where the minus sign was chosen by convenience.} \eqno (11)$$
\vskip 0.2in
\par\noindent It is not a great deal of work to find that the first derivative
at the point $\bar T= \bar T_{Rf}$ (10), is not zero and rather has the value:
\vskip 0.2in
$${d\bar T(T)\over dT}|_{\bar T_{Rf}}=-i\cosh( arcsinh \beta)\eqno
(12)$$
\vskip 0.2in
\par\noindent Therefore it is not possible to have a purely imaginary action
coming  from  the  circuit where our complex variable $T$ has a fixed real
component and a variable purely imaginary one. So we are left with the option
of the second circuit ( circuit (b) diagram above). In the first part of the
circuit (i), our variable is purely imaginary and can be rewritten as:
\vskip 0.2in
$$\bar T = iT\;;\qquad
\hbox {$T$ is real and has the following range} \qquad  0 < T < \bar
T_{If}\eqno (13)$$
\vskip 0.2in
\par\noindent Which implies that the scale factor must have the value below:
\vskip 0.2in
$$a(T)=\alpha\sin({T \over \alpha}) \eqno (14)$$
\vskip 0.2in
\par\noindent This part of the circuit gives rise to a Lorentzian metric,
contributing  with a purely imaginary action, such that our universe starts off
with a zero volume and reachs the joining surface with  a scale factor
$\alpha$. In the point $P_b$ of the diagram above, our variable turns abruptly
to be fully complex, with a fixed imaginary component and a variable real one.
There, along this straight line (ii), we could introduce the following
quantity:
\vskip 0.2in
$$\bar T = T + i\bar T_{If}\;;\qquad
\hbox {$T$ is real and has the limits} \qquad  0 < T < \bar T_{Rf} \eqno (15)$$
\vskip 0.2in
\par\noindent And from the expression of the scale factor (11) and with the
help
of (10), we can compute the value of the scale factor over this sector of our
contour:
\vskip 0.2in
$$a(T)=\alpha\cosh({T\over \alpha})\eqno (16)$$
\vskip 0.2in
\par\noindent which is real. Then this sector contributes with a Riemannian
metric and a real action. Now in this region, the universe has a scale factor
$\alpha$ on the transition surface, which increases until reachs the final
boundary surface.  As an authentic solution of our equations, its
interpretation is not a immediate process, since it points to a universe (if no
other phenomenon takes place) which is at large-scale Riemannian!

          Once the first stage of determination of the solutions of the
classical Euclidean-Einstein's equations is over, we must turn to the second
and
more complex one. We must find compact surfaces with negative curvature, which
will represent our spatial-slices. But before that it is necessary to point
out that we have arrived at a very important result, which will facilitate our
analysis of regularity immensely: the singular point in each of the  two above
solutions (7.3), (8.3) or (14), is found to be in the region where our
spacetimes are Lorentzian ones. Then, studying the regularity of one of the
solutions in this region will be enough to draw conclusion for both of them. As
a matter of simplicity, we shall choose the Lorentzian solution (7.3), which
will easier to be pictured as a whole.
\vskip 0.3in
{\bf $3$\quad -\quad Compact\quad Negatively\quad Curved\quad Surfaces\quad
in\quad Two\quad and \hfill\break \quad Three\quad Dimensions}
\vskip 0.2in
             The idea of compact surfaces with negative curvature is not
a new one [12] but its actual implementation in quantum cosmology is very
recent
[2].  It is interesting to note that in other branches of physics, e.g.
classical and quantum chaotic systems, two dimensional compact surfaces of
negative curvature  have  been  studied  for some time [13].  We  shall need
now some mathematics in order to justify the  next step, which is the
determination of the compact spatial sections of the solutions of the
Euclidean-Einstein's equations. But before that, it is important to stress
that  we  shall  start working in three spacetime dimensions in order to
understand the basic features of those compact spatial sections (in this case
two-dimensional), and also the application of our procedure to check whether
those solutions are regular or not. Then the natural next step will be the
analysis of a more realistic case in which the spacetimes will have
four-dimensions (and consequently three-dimensional compact spatial sections).

          In two dimensions, it was shown in a survey  made by Peter Scott
[14]
that every closed orientable surface of genus at least two allows a geometric
structure modeled on the two-dimensional hyperbolic space $H^2$, or in
summary admits a hyperbolic structure.  This result enables one to select an
example of a two-dimensional compact surface. It  will  be  a  closed
orientable surface of genus two.  Unfortunately, there is not such a neat
result for three-dimensional hyperbolic spaces $H^3$, but much work was done by
W. P. Thurston [15], in this area, and from his work we were able to collect
one example of a compact negatively curved surface in three-dimensions and it
will be this example, which we shall discuss below, that will represent the
spatial sections of our four-dimensional hyperbolic world.

         As we have said at the end of the last section, our analysis of
regularity will be based on the Lorentzian solution, which is simply
anti-DeSitter spacetime. Turning to Hawking \& Ellis [16], we learn that the
anti-DeSitter spacetime, can be represented as the hyperboloid:
\vskip 0.2in
$$-v^2-u^2+x^2+y^2+z^2\>=\>-1\eqno (17)$$
\vskip 0.2in
\par\noindent in the $flat$ five-dimensional space $R^5$ with metric:
\vskip 0.2in
$$ds^2\>=\>-dv^2-du^2+dx^2+dy^2+dz^2\eqno (18)$$
\vskip 0.2in
\par\noindent and our Lorentzian solution, (7.3), is obtained by the following
transformation  of variables:
\vskip 0.2in
$$\eqalignno{v\>&=\>\alpha\cos( N_I{t\over \alpha})\cr
             u\>&=\>\alpha\sin( N_I{t\over \alpha})\cosh\chi\cr
             x\>&=\>\alpha\sin( N_I{t\over \alpha})\sinh\chi\cos\theta\cr
             y\>&=\>\alpha\sin(
N_I{t\over\alpha})\sinh\chi\sin\theta\cos\phi\cr
             z\>&=\>\alpha\sin( N_I{t\over
\alpha})\sinh\chi\sin\theta\sin\phi &(19)\cr}$$
\vskip0.2in
\par\noindent It is important to comment at this point, that our solution does
not cover the whole anti-DeSitter spacetime and there is other coordinate chart
which perform this task in a more complete way [16].

          Now, if one wants to  take a glance at the spatial sections of that
manifold, one has to assign a certain value to the coordinate $t$, which is
the  same (through (19)) as specifying a value for $v$, say $v_0$. We can see
that, equation (17) is transformed to:
\vskip 0.2in
$$u^2-x^2-y^2-z^2\>=\>1-v_0^2\eqno (20)$$
\vskip 0.2in
\par\noindent And, the above equation is easily recognizable as the one for a
hyperboloid of two sheets but not in Euclidean space, rather in a Minkowskian
space. Each of the two sheets is a representation of the 3-dimensional
Hyperbolic space $H^3$, but it is very difficult to work with those surfaces in
this form.  A more profitable procedure would be to project the whole of $H^3$
into $R^3$, such that the projection is isometric. There is more than one
stereographic projection of the $H^3$, and each of these is suitable for one
certain kind of problem. At this point we shall split our analysis in two, in
order to study two and three-dimensional surfaces separately.
\vskip 0.2in
3.1\quad -\quad $2+1$\quad Spacetimes.
\vskip 0.2in
          In two dimensions, we shall use the Poincare Disc model for the
$H^2$.
The relevant spatial sections to be studied now are the ones derived by (20),
through the removal of the $z$ coordinate, or,
\vskip 0.2in
$$u^2-x^2-y^2=1-v_0^2\eqno (21)$$
\vskip 0.2in
\par\noindent This is a two-dimensional hyperboloid with two sheets embedded in
a Minkowskian space. The Poincare disc model is build up as the projection of
the upper sheet of the hyperboloid  (21) above, on the plane $(u,x,0)$, taking
as the base point of the projection, the apex $(0,0,-\sqrt{\,1-v_0^2})$. It is
important to notice, that our model differs a little from the ordinary one
[13], in the fact that the apexes are not situated at $(0,0,1)$ and $(0,0,-1)$.
The ordinary case is a limiting case when $v_0(t)$ is zero and the other
relevant limit, in fact the one we shall be concerned with is furnished when
$v_0(t)$ goes to $1$. The disc is fabricated such that its center projects the
upper apex of the hyperboloid and the outer limiting circumference of radius
$\sqrt{1-v_0^2}$, represents the points at infinity of $H^2$. There is a simple
relationship between the coordinates $(\chi, \theta)$ on the $H^2$, and the
polar coordinates $(r,\theta)$ on the P.D.,
\vskip 0.2in
$$r=\sqrt{1-v_0^2}\tanh({\chi \over 2})\>;\qquad
\hbox {$\theta$ is unchanged} \eqno (22)$$
\vskip 0.2in
\par\noindent The conformally isometric mapping produces the metric below,
\vskip 0.2in
$$ds^2=4(1-v_0^2)^2{(dr^2+r^2d\theta^2) \over (1-v_0^2-r^2)^2}\eqno (23)$$
\vskip 0.2in
          The mathematical process by which we shall obtain our compact
two-dimensional spatial sections is a tessellation [18], a covering of the P.D.
by congruent (isometric) repetition of some basic figure, the fundamental
region. The fundamental region is defined as the quotient surface $H^2/\Gamma$,
where $\Gamma$ is a certain symmetry transformation of the $H^2$.

         The relevant symmetry  transformation $\Gamma$, for a closed
orientable
region of genus two, would  be  the  ordinary translation if we were on the
Euclidean plane  ( consider the example of a torus derived from identification
of opposite sides of a square [19] ), but in the Minkowskian space the
translation  operations  are  performed  by  the Lorentzian  boosts can be
written in the linear fractional form [13]:
\vskip 0.2in
$$z'=T(z)={(\alpha z+\beta) \over (\beta^*z+\alpha^*)}\eqno (24)$$

$$\hbox {with} \qquad \vert \alpha \vert^2 + \vert \beta \vert^2 = 1 \quad
\hbox {and $z$ defined as} \quad z=re^{i\theta}$$
\vskip 0.2in
\par\noindent and $T$ has also a matrix representation given below,
\skip 0.2in
$$T=\left(\matrix{
    \alpha       &\beta       \cr
    \beta^*      &\alpha^*    \cr
    }\right)\eqno (25)$$
\vfill
\par\noindent Our compact surface will be fabricated then by the identification
of the sides, through (24), of a polygon (the relevant fundamental region) on
the P.D.. The simplest polygon is the regular octagon, with the opposite sides
identified, as shown in the figure below.
\vskip 3.0in
\centerline{3-{\sl Identified Octagon}}
\vskip 0.2in
\par\noindent For each of the four relevant directions of the transformation we
have a generator which is given in its matrix form by:
\vskip 0.2in
$$T_k=\left(\matrix{
      \cosh({\chi_0\over 2})      &e^{ik{\pi\over 4}}\sinh({\chi_0\over 2}) \cr
      e^{-ik{\pi\over 4}}\sinh({\chi_0 \over 2})      &cosh({\chi_0\over2}) \cr
      }\right)\eqno (26)$$
\vskip 0.2in
\par\noindent where $k=0,1,2,3$; and $\chi_0$ is the rapidity over which we
have
transformed the point on $H^2$. That octagonal fundamental region with
identified opposite sides, has a representation in a three-dimensional space
known as: Sphere with two handles or Double Torus ($T_2$), drawn below:
\midinsert
\vskip 1.5in
\centerline{4-{\sl Double Torus}}
\vskip 0.2in
\endinsert
\par\noindent When one notices that the metric on the P.D. is induced on the
double torus, one has arrived to the desired two-dimensional compact negatively
curved spatial sections. Then, our three-dimensional spacetime is composed of
two disjointed double tori, evolving in a symmetrical way in time.  The
analysis
made for one of them will reveal the behaviour of the other.
\vskip 0.2in
3.2\quad -\quad $3 +1$\quad Spacetimes.
\vskip 0.2in
          In this case, the model we shall use to describe the
three-dimensional
negatively curved spatial sections $H^3$,is called the projective model[15].
The first step in order to implement this projection is to repeat what was done
in the previous sub-section, to obtain the Poincare Disc model in this higher
dimensional space (P.D.3). The metric of this P.D.3 is given by the same formal
expression as (23) but now $r$ stands for the radius of a three-dimensional
spherical coordinate system. The three-dimensional Poincare Disc obtained in
this way passes through the centre of a three sphere. On this three-sphere we
can define a Northern Hemisphere and a Southern Hemisphere. The following step
is the stereographic projection of the P.D.3 on the Southern Hemisphere of the
$S^3$, taking as the base point the northern pole. Finally, perform a Euclidean
orthogonal projection of the Southern Hemisphere back to the P.D.3., and this
map of P.D.3 onto itself, along with the induced metric coming from the P.D.3,
is the projective model of $H^3$ (Pr. 3). It is important to note that in this
model the geodesics are straight lines but unfortunately the projection fails
to be conformal, in other words angles are not preserved.

           The fabrication of our desired three-dimensional compact negatively
curved spatial sections can proceed now on this projective model. Let us
start with a regular tetrahedron inscribed in the Pr.3, such that all four
vertices are on the two-sphere at infinity. The dihedral angles of this ideal
hyperbolic simplex are $\pi/3$ (here we have an astonishing verification of
the  non-conformality of the projective model). Now, let the sides and faces
expand in a homogeneous and isotropic way, until the dihedral angles reach
the value $\pi/ 6$. It can be shown, that there exists for each vertex of
our simplex, a two-dimensional surface which is perpendicular to all three
faces
having one of those vertices in common. Along those planes, truncate the
tetrahedron such that all four vertices are deleted and all points of our
simplex are inside of the Pr.3. Now, repeat all the above steps in order to get
another copy of the truncated figure $T$, then glue what was left of the former
faces following the pattern given below \hfill\break
\break\hbox{}\hfill\break
\vskip 2.5in
\centerline{5-{\sl The glueying rules to form $M$}}
\vskip 0.2in
\par\noindent The resulting complex $M$ has one boundary. One can show that
this
boundary is a two-dimensional surface of genus 2, more specifically a double
torus ( the one constructed in the previous sub-section). But we want a compact
region, without boundary. So, we must take two copies of $M$ and glue them
together by identifying the boundaries in a one-to-one map. In the following
figure we shall show the identifications of the boundaries of $T$ in order to
obtain the final compact three-dimensional surface.\hfill\break
\vfill
\break\hbox{}\hfill\break
\vskip 2.5in
\centerline{6-{\sl Identification of the boundaries of two complexes $T$}}
\vskip 0.2in
          It is that complex, which we shall call $P$,with the metric induced
by
the Pr.3 which  will  represent  each  disjoint  sequence  of three-dimensional
spatial sections of our negatively curved four-dimensional world.  As in  the
previous case it is enough to analyse the regularity of one of the sequence of
spatial sections.

             Our universe is now completed, either in  three  or
four-dimensions.
We must finally turn to the ultimate goal, the determination of the regularity
or not of those manifolds.
\vskip 0.3in
{\bf $4$\quad -\quad Regularity\quad Analysis}
\vskip 0.2in

The classical solutions for the metric ansatz have been entirely determined,
and
we  have only to answer this  question: Are they regular? Regularity conditions
have been given for spacetimes other than ours (see reference [6] for a brief
consideration of few of those proposals). Our manifolds are examples of a  very
simple class  of  solutions  of the Einstein equations: they have constant
curvature or more precisely, a constant scalar of curvature $R\;=\;-\>4\vert
\Lambda \vert$. The important point to make here is that we shall not be able
to use most of the proposals reported in reference [6], in order to identify
our spacetimes as singular or not. This is because these proposals use the
behaviour of scalar functions, derived from the curvature over the manifold, to
identify singular points. If any of these functions has an unbounded limit in a
specific point of spacetime, we say that this point is singular or has a
singularity. This is not  the  case  for  our  manifolds, for these functions
will be  constant. One would be  tempted  to  say at  this  stage  that when
all  these  scalar  functions are well behaved, it should imply that  the
manifold is regular! This would be forgetting a distinct type of singularity,
the conical singularity. Thus, even with well defined functions of the scalar
of
curvature over all their points, our manifolds could still  have a conical
singularity at the point $v_0=1$, or $t=0$.

Our analysis, will be based entirely in the application of the concept of
holonomy [20] to the manifold being studied. The holonomy group of a manifold
$M$, equipped with a metric $g$, $(M,G)$ (therefore with an affine connection
$\Gamma$), is defined as the transformations from the tangent space at  the
point  $p$, $T_pM$,  to  itself, constructed by taking a fiducial vector $v$
and  parallel  transporting it along  all closed curves starting and finishing
at $p$  on  $T_pM$.  For  a  manifold,  the set of  all  linear
transformations
generated  by  the parallel  transport  of  $v$ along all possible  closed
curves  from  $p$, is called the holonomy group of $M$ at $p$. In fact, we
shall  not  be  interested  in computing the holonomy group of  the  given
manifold  but  we  shall  use  this concept in a related way. For a manifold
$(M,g)$ and a fiducial vector we  shall  evaluate the transformation generated
by the parallel  transport  over closed curves around the  point $s$. Each of
these curves  will  have  a  certain  proper length which vanishes when $t$
tends to some limit $t_0$, as the points defined by  closed curves collapse to
the point $s$. Then, if the point $s$ is regular, the limit of the
transformation induced by the parallel transport of the fiducial vector around
a closed curve, as $t$ goes to  $t_0$,  must  be  the identity transformation.
In other words, the  limit  of  the  initial  and  final values of the vector
at $t_0$ must be the same. If this is not the case, the  manifold is
non-regular
at $s$. Let us write it down in a more precise manner.

Let $(M,g)$ be a given manifold factorized in  the  usual  $3+1$
decomposition. Let us take a point $s$ on $M$ and a certain foliation of $M$.
For $t$ being the ordinary time parameter, we choose a point $p(t_a)$ in the
neighbourhood $N(s)$ of $s$. We build up now the closed curve $c_p(t_a)$ such
that:  $c_p(t_a)\>=\>[c(\lambda,t_a)\>/\> c(\lambda_0,t_a)=
p(t_a)\>=\>c(\lambda_f,t_a)\;;\;\lambda_0<\lambda<\lambda_f]$. We furthermore
imply that: for $p(t_b) \in N(s)$, if $t_a \not= t_b$ then, $p(t_b) \not\in
c_p(t_a)$. With the curves $c_p(t)$ we define the map $T_p(t_a)$:\quad $p(t_a)
\mapsto p(t_a)$ / it is $\nabla_{u(c_p(t_a))}v$ around $c_p(t_a)$.For each
$c_p(t)$, we define a function $L$ such that $L(c_p(t))$ gives one possible
type of length. Now we introduce $t_0$, which is defined by the limit when
$L(c_p(t))$ vanishes and $p(t_a)$ goes to $s$. We require for $M$ to be
regular, the limit $\lim_{t \to t_0} T_p(t_a)$ to be the identity. Now that
our  method  of analysis is well established, let us turn to  the  specific
manifolds  of  interest.
\vskip 0.2in
4.1\quad -\quad  $2 +1$ \quad Spacetimes.
\vskip 0.2in
The first solution to be analysed and also the one we shall consider in
deepest details will be the one with three spacetime dimensions and two-
dimensional spatial sections, given in $3.1$. The manifold $M$ is defined for
this case, by the metric (7.3) without the component $g_{\phi \phi}$,  and the
two-dimensional metric of the compact spatial foliation derived in (23) with
the aid of the coordinate transformation (22) will be used. We shall also
change
our coordinates in (23), from polar to  Cartesian (which will simplify our
study) and finally set $N_I$ and $\alpha$ to one. The starting point is the
following metric,
\vskip 0.2in
$$ds^2\>=\>-dt^2+{4(1-v_0)^2(dx^2+dy^2) \over (1-v_0^2-x^2-y^2)^2}\eqno (27)$$
\vskip 0.2in
\par\noindent with the simplifications we made, $v_0$ is reduced to $\cos t$,
from (19). The fact that the manifold under study is a Lorentzian one will be
helpful, since for this type of spacetime, no other information than the above
mentioned limit of the map $T_p(t)$ is required in order to establish its
regularity.
\par\noindent $------------------------------------$
\par\noindent As pointed out by J. Louko in a private communication, for
Riemannian manifolds one can obtain unity limit for the map $T_p(t)$ even for
singular points on the manifold. Then, it would be important to have some
extra information about the functions composing the metric.
\par\noindent $------------------------------------$
\par\noindent The next step is to find out the transformation $T_p(t)$, which
will be possible only when we have the parallel transport equations. The
easiest way to derive them from (27) is by working on the orthonormal basis,
defined by the transformation: \vskip 0.2in $$w^t = dt\;;$$ $$w^i =
{2(1-v_0^2)dx_i \over (1-v_0^2-x^2-y^2)}\quad \hbox {for} \quad x_1=x\quad
\hbox {and} \quad x_2=y \eqno (28)$$ \vskip 0.2in
\par\noindent The non-vanishing connection coefficients components in this base
are:
\vskip 0.2in
$$\Gamma^t_{xx}\>=\>{2\dot
v_0(t)v_0(x^2+y^2) \over (1-v_0^2)(1-v_0^2-x^2-y^2)}\>=\>\Gamma^x_{tx}\eqno
(29)$$
$$\Gamma^t_{yy}\>=\>{2\dot
v_0(t)v_0(x^2+y^2) \over (1-v_0^2)(1-v_0^2-x^2-y^2)}\>=\>\Gamma^y_{ty}\eqno
(30)$$
$$\Gamma^x_{yy}\>=\>{-x \over 1-v_0^2}\>=\>-\Gamma^y_{xy}\eqno (31)$$
$$\Gamma^y_{xx}\>=\>{-y \over 1-v_0^2}\>=\>-\Gamma^x_{yx}\eqno (32)$$
\vskip 0.2in
\par\noindent In this base, the general expression for the parallel transport
equation is [21]:
\vskip 0.2in
$${dv^\alpha \over
d\lambda}\,+\,v^\beta \Gamma^\alpha_{\beta \gamma}{dx^\gamma \over
d\lambda}\,=\, 0 \eqno (33)$$
\vskip 0.2in
\par\noindent Two important remarks must be make at this point. The first is
that, as we  stressed before, the map $T_p(t)$ is computed for one given value
of the coordinate $t$. It means that at the set of equations (33), we must
consider the path or paths for constant $t$. The other point is related to the
fact that we are  working with a non-coordinate basis and in reality we are
willing to describe our paths in terms of elements of a coordinate basis. In
order to do this we shall have to modify our set of parallel transport
equations and introduce the relevant transformation matrix. Then, the new set
of equation is written,
\vskip 0.2in
$${dv^\alpha \over d\lambda}\,+\,
v^\beta\Gamma^\alpha_{\beta\gamma}\Lambda^\gamma_\delta {dx^\delta \over
d\lambda}\,=\,0 \eqno (34)$$
\vskip 0.2in
\par\noindent where $\Lambda$ is the basis one-form transformation matrix, also
responsible  for the transformation of vector components. We must now
concentrate on the following question: what are all possible independent closed
curves defined  at one moment of time in our compact two-dimensional spatial
surface? The answer comes from geometrical considerations on our fundamental
region, figure $3$. For our closed paths to take in account the fact that our
fundamental region is compact, they should use the identification of the
opposite sides at least once. In other words, our paths must exit from one side
and reappear in the opposite one, at least one time. But of course, the paths
can do so more than once, which leave us with a infinity number of closed
trajectories to be analysed. Fortunately, that infinity number is only apparent
because they are not all independent paths, more than that, they are all
constructed out of a finite number of paths. The basic paths are in reality
just four ( and also the ones derived from those basic ones by means of allowed
symmetry transformations, as parity and rotation under some specific angles)
and
they are the only possible ones which exit the fundamental region at one side
reappearing at the opposite side once and just once.  So, all other paths which
use the identification of opposite sides more than once are built up as
combinations of those basic ones, which are given below:
\vfill
\break\hbox{}\hfil\break
\vskip 3.0in
\centerline{7-{\sl Basic loops on the identified octagon}}
\vfill
Then, in order to demonstrate the regularity of our manifold, we should
determine the transformations $T_p(t)$,and its limit as $v_0$ goes to $1$ or
identically as $t$ goes to $0$, over all those four closed curves. On the other
hand, if we find out that for one of the four $T_p(t)$ defined in that way, we
do not have the desired limit in order for $M$  to be a regular manifold,  we
shall be able to conclude without any more argument that our manifold is
not regular at the spacetime event under consideration. Let us start with the
simplest one, (a), at the figure 7 above.

This path, is defined by: $y=0$ and we can set the coordinate $x$ to be
described by the simplest non-constant function of the parameter $\lambda$,
which is $x\,=\,\lambda$ with the resulting range for $\lambda:\quad -x_0(t) <
\lambda < x_0(t)$. It is important to note here, that, as $\lambda$ is not a
function of $t$, all information about the contraction or expansion of our
fundamental region, as the time  passes is contained at the extreme values of
$\lambda$, which are given by (22):
\vskip 0.2in
$$x\,=\,\sqrt{1-v_0^2}\tanh({\chi_0 \over 2})\quad;\quad
\theta\>=\>0,\pi \eqno (35)$$
\vskip 0.2in
\par\noindent where $\chi_0$ is one fixed value of $\chi$. Then the parallel
transport equations (34),  becomes, with the aid of (28) (for the relevant
components of the basis one-form transformation $\Lambda$), (29), (31) and (32)
(for the necessary connection coefficient components):
\vskip 0.2in
$${dv^y \over d\lambda}\,=\,0 \eqno (36)$$
$${dv^t \over d\eta} + {2\beta\eta^2v^x \over \alpha(1-\eta^2)^2}\,=\,0
\eqno (37)$$
$${dv^x \over d\eta} + {2\beta\eta^2v^t \over \alpha(1-\eta^2)^2}\,=\,0
\eqno (38)$$
\vskip 0.2in
\par\noindent where we have defined:
\vskip 0.2in
$$\beta\,=\,2\dot v_0v_0 \quad;\quad \alpha^2\,=\,1-v_0^2 \eqno (39)$$
\vskip 0.2in
\par\noindent and we also have defined the new variable $\eta$,  related  to
$\lambda$  by, $\eta\,=\,\lambda / \alpha$. The range of this new variable is
given according to  its  definition by: $-x_0(t)/ \alpha < \eta < x_0(t)/
\alpha$.  But  from  (35)  we  see  that  the limits are simply $\pm
\tanh(\chi_0/2)$, which  are  independent  of  $t$,  then  all  the dependence
of $T_p(t)$ now must come exclusively from the quantities $\alpha$ and $\beta$
defined above.

The solution of the equation (36) is straightforward, and for a initial
value of the component y of the fiducial vector to be parallel transported,
equal to $v^y_0$ (or $v^y(\eta=-\eta_0)=v^y_0$), we have,
\vskip 0.2in
$$v^y\,=\,v^y_0 \eqno (40)$$
\vskip 0.2in
\par\noindent The other two equations form a coupled system of first order
differential  equations and its solution will require a very particular change
of variable as we shall see next. Define a new variable $\rho$ such that the
differential variation of it is related to the differential variation of the
old one $\lambda$ by: $d\rho\,=\,2\beta\eta^2/\alpha(1-\eta^2)^2$. Through
explicit integration we can obtain a relationship between the variables
themselves:
\vskip 0.2in
$$\rho\,=\,{\beta \over \alpha}[{\eta \over
1-\eta^2}-{1 \over 2}\log({1+\eta \over 1-\eta})] \eqno (41)$$
\vskip 0.2in
\par\noindent and the range of this new variable is given by:
\vskip 0.2in
$$-\rho_0\,<\,\rho\,<\,\rho_0 \eqno (42)$$
\vskip 0.2in
\par\noindent where $\rho_0$ is defined such that
$\rho(\eta=\eta_0)\,=\,\rho_0$
and $\rho(\eta=-\eta_0)$ is  easily seen to be $-\rho_0$. The equations (37)
and
(38) are now expressed in  terms of $\rho$ in the following very pure way:
\vskip 0.2in
$${dv^t \over d\rho} + v^x\,=\,0 \eqno (43)$$
$${dv^x \over d\rho} + v^t\,=\,0 \eqno (44)$$
\vskip 0.2in
\par\noindent The general solution is given in terms of two constants, A and
$\delta_0$, to be determined by the initial conditions, and that solution is:
\vskip 0.2in
$$v^x(\rho)\,=\,A\sinh(\rho+\delta_0) \eqno (45)$$
$$v^t(\rho)\,=\,A\cosh(\rho+\delta_0) \eqno (46)$$
\vskip 0.2in
\par\noindent The initial conditions will  be given by:
$v^x(\rho=-\rho_0)\,=\,v^x_0$ and $v^t(\rho=-\rho_0)\,=\,v^t_0$. If one
introduces those quantities in the equations (45) and (46), after some
algebraic
calculations one finds,
\vskip 0.2in
$$v^x(\rho)\,=\,\cosh(\rho+\rho_0)v^x_0+\sinh(\rho+\rho_0)v^t_0 \eqno (47)$$
$$v^t(\rho)\,=\,\sinh(\rho+\rho_0)v^x_0+\cosh(\rho+\rho_0)v^t_0 \eqno (48)$$
\vskip 0.2in
\par\noindent Then, joining together all the solutions of our parallel
transport
equations, (40), (47) and (48), we can write them in a matrix form, from
which it will be straightforward for one to read the $T_p(t)$ map.
\vskip 0.2in
$$v(\rho)\>=\>T_p(t)v_0,\qquad \hbox {where}$$
\vskip 0.2in
$$T_p(t)=\left(\matrix{
         \cosh(\rho+\rho_0)  &\sinh(\rho+\rho_0)  &0  \cr
         \sinh(\rho+\rho_0)  &\cosh(\rho+\rho_0)  &0  \cr
          0                  &0                   &1  \cr
         }\right) \eqno (49)$$
\vskip 0.2in
\par\noindent For simplicity and relaxing a little the proper use of the
concepts,  we  shall refer to the matrix $T_p(t)$ as "holonomy matrix". It is
important to notice  the general behaviour of a fiducial vector parallel
transported  along  the  chosen closed curve, which can be readily seen with
the
help of the holonomy  matrix. If we set $\rho=-\rho_0$, at (49), which means
that we are in the initial point, we get the initial value of the fiducial
vector, as  defined.  But  if  we  set  $\rho=\rho_0$, in general, our vector
does not return to the initial  value,  although  by definition our curve has
returned to its initial point. This fact, was expected, and it is just a
verification that our manifold has a non-vanishing  curvature. The final step
in our present analysis, is the evaluation of the holonomy matrix after a
complete loop in our closed curve in the limit when $t$ goes to $0$.

The relevant components of the holonomy matrix to be analysed when
we have performed a complete loop around the path, or in other words at the
spacetime point $\rho=\rho_0$ are:
\vskip 0.2in
$$\cosh(2\rho_0)\qquad \hbox {and} \qquad \sinh(2\rho_0) \eqno (50)$$
\vskip 0.2in
\par\noindent We should now take the limit of those quantities when t goes to
zero. The starting point must be the explicit expressions of (50), in terms of
the variable time.  This can be achieved when one notice that from the
definition of the terms $\alpha$ and $\beta$ (39), and the fact already
mentioned that $v_0$ is simply $\cos t$ we obtain the following value for the
ratio $\beta/\alpha$:
\vskip 0.2in
$${\beta \over \alpha}\,=\,-2\cos t \eqno (51)$$
\vskip 0.2in
\par\noindent This ratio is where all dependence on t of the functions (50) is
concentrated. It is readily seen by direct observation of the expression (41),
which gives the relationship between the variables $\rho$ and $\eta$, when we
set $\rho\,=\,\rho_0$. When one takes the exponential of both sides of this
expression evaluated at that particular value of $\rho$, or better saying
$\eta$, one obtains:
\vskip 0.2in
$$e^{2\rho_0}\,=\,({1+\tanh ({\chi_0
\over 2}) \over 1-\tanh ({\chi_0
\over 2})})^{2\cos t}e^{-4\cos t\sinh (\chi_0)} \eqno (52)$$
\vskip 0.2in
\par\noindent That expression and the one derived from it by replacing $\rho$
by
$-\rho$, are the  building blocks of the quantities given in (50). Then, the
limit of the desired elements of the holonomy matrix, (50), when $t$ goes to
$0$ is:
\vskip 0.2in
$$\lim_{t \to 0}\cosh2\rho_0\,=\,\cosh[2(\chi_0-\sinh(\chi_0))] \eqno (53)$$
$$\lim_{t \to 0}\sinh2\rho_0\,=\,\sinh[2(\chi_0-\sinh(\chi_0))] \eqno (54)$$
\vskip 0.2in
\par\noindent It is not difficult to see that the holonomy matrix will go to
the
unity at that limit, if and only if, the arguments of the hyperbolic functions
on the right-hand side of the expressions (53) and (54) vanish. The unique
possibility for this to happen, is if the quantity $\chi_0$, which gives the
initial size of our fundamental region is zero. But it is not possible since
our
fundamental region starts with a finite, non-vanishing well defined size. We
are then forced to conclude that there is a singularity at the point $t=0$, of
our manifold, for this choice of closed path. But, if for one of the four basic
closed loops of our fundamental region, we found that there is a singularity
sat at the point $t=0$, it must not be a particularity of the chosen  closed
path. Then, irrespective of the closed basic curve we chose we should find the
same result. More than that, this result is also independent of the coordinates
in which one decides to describe our spacetime, as long as, one keeps untouched
one of the basic features of our anti-DeSitter model, the homogeneity of its
spatial sections. It is more clearly appreciated if one realizes that the
effect of a basis one-form coordinate transformation of the metric components,
given by the S matrix, upon our holonomy matrix is the similarity
transformation
$ST_p(t)S^{-1}$.But it is easy to see that no transformation of this kind would
be able to furnish the desired unity limit to our holonomy matrix, when $t$
goes
to $0$. So, we must finally answer the question posed by the title of this
article: Yes, the 3-dimensional pseudo-sphere universe had a beginning.
\vskip 0.2in
4.2\quad-\quad $3 +1$\quad Spacetimes.
\vskip 0.2in
The results and the various steps followed in the previous section,
will form the core of the work planned for the present section. We shall be
interested in compute the holonomy matrix for the more physical 4-dimensional
version of the anti-DeSitter spacetime given by the Lorentzian solution (7.3),
now in its complete form, and the three-dimensional compact spatial sections
will be described by the three-dimensional version of (23), in Cartesian
coordinates, when the coordinate transformation (22) holds. The explicit form
of  the metric is then, \vskip 0.2in
$$ds^2\,=\,-dt^2+{4(1-v_0^2)^2(dx^2+dy^2+dz^2) \over
(1-v_0^2-x^2-y^2-z^2)^2} \eqno (55)$$
\vskip 0.2in
\par\noindent where the same simplifications for $\alpha$ and $N_I$ were made,
leading to a $v_0\,=\,\cos t$. The non-vanishing components of the connection
coefficients for that metric (55) are:
\vskip 0.2in
$$\Gamma^t_{ii}\,=\,{2\dot
v_0v_0(x^2+y^2+z^2) \over (1-v_0^2)(1-v_0^2-x^2-y^2-z^2)}\,=\,\Gamma^i_{ti}$$
$$\Gamma^i_{jj}={-i \over 1-v_0^2}\,=\,-\Gamma^j_{ij}\quad \hbox {where}
\quad i,j\,=\,x,y \& z \eqno (57)$$
\vskip 0.2in
{}From here onward the analysis will be a repetition of the previous
case, except by the choice of the basic closed loops on our three-dimensional
fundamental region given now by the complex $P$, introduced in 3.2. As one
can  imagine, it will be much more difficult to find in this case the analogue
of the four basic closed curves on the identified octagon. Because of this huge
complexity and also due to the fact that we have learned how to derive the
limit of $T_p(t)$ out of a very simple closed loop, in the last section, we
shall not try to identify all possible basic closed paths on the complex $P$.
Rather, we shall try to find the path similar to the one we used on the section
4.1, on the identified octagon on our present fundamental region. The desired
closed curve is not difficult to be found, one has only to inspect two
complexes $T$, figure 6, disposed in such a way that one out of the  five
one-to-one identification between two of the ten boundaries (five for  each
complex) is displayed explicitly. This can be seen in the figure below:
\vfill
\break\hbox{}\hfil\break
\vskip 3.0in
\centerline{ 8-{\sl Basic loop in the fundamental complex P}}
\vskip 0.3in
It is easy to see now, that with the Cartesian coordinate axis placed as
in the figure above, the curve C' which runs along with the $z$ axis, starting
at $-z_0(t)$ and moving up until $z_0(t)$, and of course returning to
$-z_0(t)$,
since both boundaries at $z_0(t)$ and $-z_0(t)$ are identified, is the analogue
of the basic loop we chose in the last section. Then, chosing this closed
path to parallel transport our fiducial vector we shall be able to determine
like before a holonomy matrix and finally its regularity as $t$ goes to $0$.
Introducing the equation for the curve C':
\vskip 0.2in
$$x\,=\,y\,=\,0\quad \hbox {and} \quad z\,=\,\lambda\,;\qquad \hbox {where}
\qquad -z_0(t)\> < \>\lambda\> <\> z_0(t) \eqno (57)$$
\vskip 0.2in
\par\noindent on the parallel transport equations (34), (and the basis one-form
transformation (28) suitably modified to deal with the presence of an extra
dimensional coordinate) along with the non-vanishing components of the
connection coefficients (56), we get for the initial conditions:
\vskip 0.2in
$$v^t(\lambda=-z_0(t))\,=\,v^t_0 \qquad ;\qquad
v^x(\lambda=-z_0(t))\,=\,v^t_0 \eqno (58)$$
$$v^y(\lambda=-z_0(t))\,=\,v^y_0 \qquad ;\qquad
v^z(\lambda=-z_0(t))\,=\,v^z_0 \eqno (59)$$
\vskip 0.2in
\par\noindent The following holonomy matrix in terms of the variable $\rho$
introduced in the last section:
\vskip 0.2in
$$v(\rho)\;=\;T_p(t)v_0\qquad \hbox {where} \eqno (60)$$
\vskip 0.2in
$$T_p(t) = \left(\matrix{
           \cosh(\rho+\rho_0)  &0  &0  &\sinh(\rho+\rho_0)  \cr
             0                 &1  &0  &0                   \cr
             0                 &0  &1  &0                   \cr
           \sinh(\rho+\rho_0)  &0  &0  &\cosh(\rho+\rho_0)  \cr
           }\right) \eqno (61)$$
\vskip 0.2in
\par\noindent and the limit of the relevant elements of $T_p(t)$ given above,
when $t$ goes to $0$, is given by (53) and (54). And as before, we conclude
this time for the four-dimensional version of the anti-DeSitter space time
equipped with compact spatial sections that there is a singularity placed at
$t=0$.
\vskip 0.3in
{\bf V\quad -\quad Conclusions}

          The results obtained in this article prompt some questions: Do all
negatively curved spacetimes decompose, after a deformation of the contour
of integration on the complex $T$ plane (figure 1), into an initial Lorentzian
spacetime and a later Riemannian spacetime? Are those space-times always
singular, preventing then the construction of no-boundary wave-function?  We
are working with a simple model and cannot derive such a general result.

          Another line of work is suggested by the conclusions we have got
here.
Supposing it turns out that the non-regularity of negatively curved Lorentzian
spacetimes is a general one, and one wants to try to construct a sensible
wave-function for those manifolds. It is clear that the initial conditions will
have to take in account the presence of the singularity. But how? The first and
simplest idea is to add to the empty spacetime a matter distribution which
could conceal out the singular behaviour of the given universe. Is there any
matter distribution which gives rise to a conical like singularity? In $2+1$
dimensions, S. Deser and R. Jackiw [22], have shown that a negatively curved
spacetime like ours, where all the curvature comes from the presence of a
cosmological constant, develop  conical-like singularities around point-like
mass-energy distributions. For a $3+1$ dimensional universe, the analogy with
the $2+1$ dimensional case [22], [23] and some results for $flat$
four-dimensional spacetimes [24], imply that the most natural candidate for a
mass-energy distribution generating a smooth spacetime is an infinite string.
Then, one would be lead to include scalar matter fields as well as gravity.
\vskip 0.3in
\centerline {{\bf References:}}

\noindent \item{[1]} Y. Fujiwara et al. 1990 {\it Phys. Rev.} D {\bf 44} 1756
\item {} Y. Fujiwara et al. 1991 {\it Phys. Rev.} D {\bf 44} 1763
\item {} Y. Fujiwara et al. 1992 {\it Class. Quantum Grav.} {\bf 7} 163
\item {} S. Carlip 1992 {\it Phys. Rev.} D {\bf 46} 4387

\noindent \item {[2]} H. P. Nilles 1984 {\it Phys. Rep.} {\bf 110} 1-162
\item {} P. P. Srivastava 1986 {\it Supersymmetry, Superfields and
Supergravity:
An Introduction} (Bristol and Boston: Adam Hilger in association with The
University of Sussex Press) 151-152

\noindent \item {[3]} J. A. Wheeler 1968 {\it Batelles Rencontres} eds C.
DeWitt
and J. A. Wheeler (New York: Benjamin) 242

\noindent \item {[4]} J. B. Hartle and S. W. Hawking 1983 {\it Phys. Rev.} D
{\bf 28} 2960
\item {} J. Louko and P. J. Ruback 1991 {\it Class. Quantum Grav.} {\bf 8} 91

\noindent \item {[5]} The first reference in [4]

\noindent \item {[6]} R. M. Wald 1984 {\it General Relativity} (Chicago and
London: The Univ. of Chicago Press) 95 and 211

\noindent \item {[7]} G. W. Gibbons and J. B. Hartle 1990 {\it Phys. Rev.} D
{\bf
42} 2458

\noindent \item {[8]} J. J. Halliwell 1991 {\it Proceedings of the Jerusalem
Winter School on Quantum Cosmology and Baby Universes} eds S. Coleman, J. B.
Hartle, T. Piran and S. Weinberg (Singapore:World Scientific) 65-157

\noindent \item {[9]} J. J. Halliwell and J. Louko 1990 {\it Phys. Rev.} D
{\bf 42} 3997

\noindent \item {[10]} J.J. Halliwell and J. B. Hartle 1990 {\it Phys. Rev.} D
{\bf 41} 1815

\noindent \item {[12]} J. L. Anderson 1967 {\it Principles of Relativity
Physics}
(New York and London: Academic Press) 454-455

\noindent \item {[13]} N. L. Balazs and  A. Voros 1986 {\it Phys. Rep.} {\bf
143} 109-240

\noindent \item {[14]} P. Scott 1983 {\it Bull. London Math. Soc.} {\bf 15}
401-487.

\noindent \item {[15]} W. P. Thurston 1982 {\it The Geometry and Topology of
Three-Manifolds} (Princeton: Princeton University Press) 3.6-3.9

\noindent \item {[16]} S. W. Hawking and G. F. R. Ellis 1973 {\it The Large
Scale Structure of Space-Time} (Cambridge: Cambridge University Press) 124-134

\noindent \item {[17]} H. S. M. Coxeter 1969 {\it Introduction to Geometry}
(New
York: John Wiley \& Sons) 287-304
\item {} A. F. Beardon 1983 {\it The Geometry of Discrete Groups (Graduate
Texts
in Mathematics)} (New York: Springer-Verlag) 126-187

\noindent \item {[18]} C. Series 1987 {\it Dynamical Chaos - Proceedings of a
Royal Society Discussion Meeting} eds M. V. Berry, I. C. Percival and N. O.
Weiss  (Princeton: Princeton University Press) 171

\noindent \item {[19]} M. Nakahara 1990 {\it Geometry, Topology and Physics
(Graduate Student Series in Physics)} (Bristol and New York: Adam Hilger)
38-42

\noindent \item {[20]} For an introductory approach see [19] pages 232-234 and
M. Crampin and F. A. E. Pirani {\it Applicable Differential Geometry (London
Mathematical Society Lecture Note Series 59)}  (Cambridge: Cambridge University
Press) 378
\item {} For a more advanced consideration, see S. Salamon 1989 {\it
Riemannian Geometry and Holonomy Groups (Pitman Research Notes in Mathematics
Series 201)} (New York: Longman Scientific \& Technical)

\noindent \item {[21]} C. W. Misner, K. S. Thorne, J. A. Wheeler 1973
{\it Gravitation} (New York: W. H. Freeman and Company) 245-263

\noindent \item {[22]} S. Deser and R. Jackiw 1984 {\it Ann. of Phys.} {\bf
153}
405

\noindent \item {[23]} S. Deser, R. Jackiw and G. t'Hooft 1984 {\it Ann. Phys.}
{\bf 152} 220

\noindent \item {[24]} A. Vilenkin 1981 {\it Phys. Rev.} D {\bf 23} 852
\item {} A. Vilenkin 1985 {\it Phys. Rep.} {\bf 121} 263
\item {} A. Vilenkin 1987 {\it Three Hundred Years of Gravitation} eds S. W.
Hawking and W. Israel (Cambridge: Cambridge University Press) 499-523
\item {} W. A. Hiscock 1985 {\it Phys. Rev.} D {\bf 31} 3288
\item {} D. Garfinkle 1985 {\it Phys. Rev.} D {\bf 32} 1323
\vskip 0.3in
\centerline {{\bf Acknowledgements}}
\vskip 0.4in
I am grateful to I.G. Moss for suggestive discussions in the course of this
work. I would like also to thanks G. Huish for advice on the two most recent
articles on reference [1], I. Raptis for advice on reference [13] and for
some intense discussions, J. Barrett \& P. Parker for advice on reference [14]
and J. Louko for his private comment on regularity of Riemannian Manifolds.
Finally I would like to thank CAPES of Brazil for the invaluable financial
support.
\vskip 0.3in
\centerline {{\bf Figure Captions}}
\vskip 0.4in

Figure 1 - Tunneling Solution

Figure 2 - Deformations of contour on the complex $\bar T$ plane

Figure 3 - Identified Octagon

Figure 4 - Double Torus

Figure 5 - The glueying rules to form $M$

Figure 6 - Identification of the boundaries of two complexes $T$

Figure 7 - Basic loops on the identified octagon

Figure 8 - Basic loop in the fundamental complex $P$
\end